\documentstyle[multicol,aps,epsf]{revtex}

\begin{document}

\title{Conductance Oscillations in Transition Metal Superlattices}
\author{S. Sanvito\thanks{e-mail: sanvito@dera.gov.uk},
C.J. Lambert\thanks{e-mail:c.lambert@lancaster.ac.uk}}
\address{School of Physics and Chemistry, Lancaster University,
Lancaster, LA1 4YB, UK}
\author{J.H. Jefferson,}
\address{Defence Evaluation and Research Agency,  Electronics Sector,
Malvern, Worcs. WR14 3PS UK}
\author{A.M. Bratkovsky,}
\address{Hewlett Packard Laboratories, Palo Alto, CA94304-1392}
\date{\today}
\maketitle
\begin{abstract}
We present a numerical study of conductance oscillations of transition
metal multilayers as a function of layer thickness. Using a material-specific
tight-binding model, we show
that for disorder-free layers
with random thicknesses but clean interfaces, long-period oscillations
in the conductance can occur, which are reminiscent of those found
in structures exhibiting GMR.
Using a heuristic effective mass model, we argue that these oscillations
arise from beating between the Fermi wavevector and a class of wavevectors
characteristic of the superlattice structure.
\end{abstract}

\begin{multicols}{2}


Oscillations in transport properties of metallic superlattices
have been largely studied in magnetic/non-magnetic multilayers, which exhibit
giant magnetoresistance (GMR) \cite{bab,barnas,prat}. In such systems
the magnetic configuration
of the superlattices is an oscillating function of the layer
thickness, switching periodically between ferromagnetic and antiferromagnetic
alignment of adjacent magnetic layers, with typical periods extending
over several atomic planes.
This arises from an oscillating inter-layer exchange coupling and results
in oscillations of the overall conductance, with the antiferromagnetic
configuration being the more resistive.

Recently a new set of measurements on  Ni/Co \cite{gall1,gall2,12bis},
Ag/Pd \cite{kim}, Ag/Au and Ag/Cu \cite{12bis} multilayers revealed
the possibility of long-period oscillations of a different origin.
On the one hand, the Ag based multilayers are entirely non-magnetic.
On the other,
the Ni/Co multilayers were measured in high magnetic field, far above the
coercive field of the structure, which rules out  magnetic
misalignment between magnetic layers as source of the oscillations.
In these experiments, all the measurements were conducted with the
current in plane (CIP) configuration and the
authors associated the oscillations 
with the  formation of a d superlattice bound state, giving rise
to a strong s-d scattering.
 
In this letter we predict that
such oscillating behaviour can also occur
with the current perpendicular to the planes (CPP),
in clean superlattices with very good interfaces, but where
the layer thickness fluctuates randomly. To the best of our knowledge
this is the first time that conductance oscillations in the CPP configuration
have been identified.

\begin{figure}
\narrowtext
\epsfysize=5cm
\epsfxsize=7cm
\centerline{\epsffile{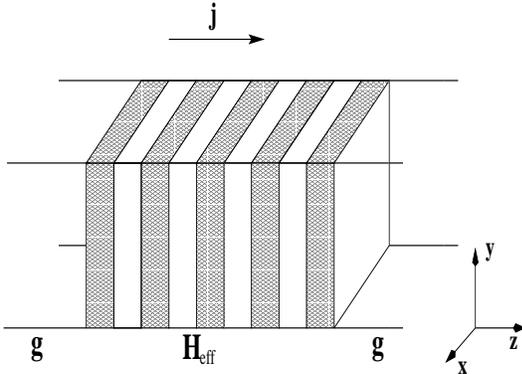}}
\caption{ \label{structure}
Finite multilayer connected to pure crystalline seminfinite leads. $g$ are the
surface Green function describing the leads and $H_{\mathrm eff}$ is the effective
hamiltonian describing the multilayer.}
\end{figure}

To address this problem, we have developed a very
efficient technique to calculate  transport properties
of a finite multilayer attached to semi-infinite pure crystalline leads, as
sketched in figure \ref{structure}. Our
calculations are based on the Landauer-B\"uttiker formalism \cite{but},
using {\it ab initio} s-p-d tight-binding Hamiltonians with nearest-neighbor
hopping.
The conductance $\Gamma^\sigma$ of a given spin species
is obtained from the Landauer formula \cite{but}
\begin{equation} \label{lanfor}
    \Gamma^\sigma=\frac{e^2}{h}T^\sigma,
\end{equation}
where $T^\sigma$ is the total transmission coefficient for the spin $\sigma$
($\sigma=\uparrow,\downarrow$)
calculated at the Fermi energy. The latter is obtained by extracting the
$S$ matrix from the total Green function $G$ of the superlattice in contact with external
leads. The total Green function
is calculated via Dyson's
equation, starting from the surface Green function $g$ of the leads and an
effective Hamiltonian $H_{\mathrm eff}$ describing the finite multilayer. 
A detailed description of the
technique will be given elsewhere \cite{noi}. In all the following calculations we
consider a perfect lattice match between clean fcc layers and hence
$k_\parallel$ is a good quantum number (the symbol $\parallel$ represents
the in-plane coordinates and the symbol $\perp$ the direction of the current
perpendicular to the planes).
The Hamiltonian is diagonalised in the Bloch basis to yield
\begin{equation} \label{lanblo}
    \Gamma^\sigma=\sum_{k_\parallel}\Gamma^\sigma(k_\parallel)=
    \frac{e^2}{h}\sum_{k_\parallel}T^\sigma(k_\parallel),
\end{equation}
where the sum over $k_\parallel$  extends over the two dimensional
Brillouin zone.
In what follows,
we employ of order
$10^4$  $k_\parallel$-points, which is sufficient
to render effects due to the finite
number of $k_\parallel$-points negligible compared with
the oscillations of interest.
In what follows, for Ni/Co and Ag/Pd multilayers,
we calculate the total conductance of the two independent spin channels
as a function of layer thickness,
in the limit that the spin-flip and phase-breaking
lengths are infinite.
It should be noted that the majority bands of Ni and Co
are s-p-like and are closely aligned. On the other hand
the minority bands are d-like and possess a relative shift in energy of about
0.7 eV. Hence we expect a large contribution to the conductance
from the majority channel and a small contribution from
the minority channel. For Ag/Pd the situation is qualitatively
different, because at the Fermi energy the DOS of Ag
is dominated by s-p electrons, while in Pd it is dominated by  d electrons.
As a consequence one expects strong interband scattering at the
interfaces between the different metals.

Following reference \cite{matto}, we
consider a pseudorandom layer arrangement, in which a finite A/B
multilayer, attached to semi-infinite leads of material
A, possesses B-layers of fixed thickness $l_{\mathrm B}$
and  A-layers of random thicknesses $l_{\mathrm A}$
which are allowed to
fluctuate by $\pm 1$ atomic planes (AP) around a mean value
$\bar{l}_{\mathrm A}$. In all the following simulations, we consider
multilayers consisting of 10 A/B bilayers and
for each $l_{\mathrm B}$, show results for the average conductance of
10 random configurations of the A-layers.

\begin{figure}
\narrowtext
\epsfysize=6cm
\epsfxsize=7cm
\centerline{\epsffile{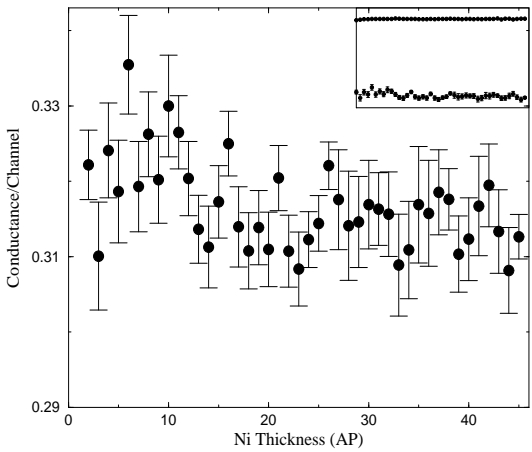}}
\caption{ \label{nico}
Conductance of Ni/Co multilayers as a function of the Ni
thickness. 
The Co thickness is 10 atomic planes.
The inset shows the two spin conductances on the same
scale with the upper plot for majority spin and 
the lower for minority spin.}
\end{figure}

\begin{figure}
\narrowtext
\epsfysize=6cm
\epsfxsize=7cm
\centerline{\epsffile{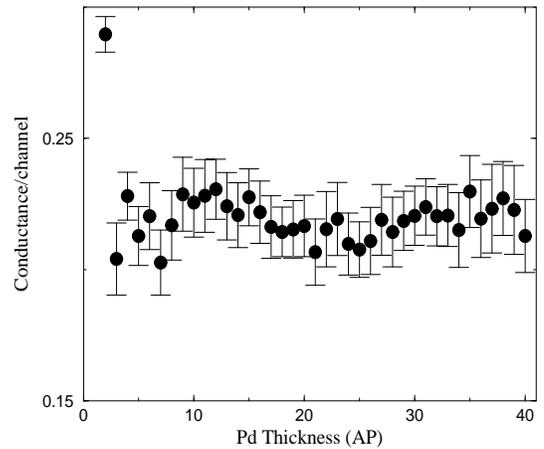}}
\caption{ \label{agpd}
Conductance of Ag/Pd multilayers as a function of the Pd thickness with
an average Ag thickness of 5 atomic planes.}
\end{figure}

For Ni/Co and Ag/Pd respectively,
figures \ref{nico} and \ref{agpd} show
the mean conductance as a function of $l_{\mathrm B}$, along
with the corresponding root-mean-square deviation from the mean.
These figures demonstrate the presence of
long-period oscillations,
with amplitudes  not exceeding $25\%$ of the mean conductance.
Moreover
the Ni/Co system shows smaller oscillations than the Ag/Pd system, and despite
the fact that the conductance of the majority spin channel is almost double that
of the minority, the
oscillations arise predominantly from the minority spins, where the scattering
is strongest.
Bearing in mind
that these calculations involve the CPP configuration, the size of the
oscillations compared with the mean conductance for the Ni/Co system,
is consistent with some of the experimental data \cite{12bis}. 

To understand how quantum interference of the conduction
electron wave-functions can give rise to such long period oscillations,
we now develop a heuristic continuum model,
within the effective mass approximation, describing
an infinite 3D superlattice with a
Kronig-Penney potential and a parabolic band.
The spin-dependent Hamiltonian for such
a system is
\begin{equation} \label{kph}
    H^\sigma({\bf r})=-\frac{\hbar^2}{2}\left[\frac{\nabla^2_{xy}}{m^*(z)}+
    \frac{\partial}{\partial z}\frac{1}{m^*(z)}\frac{\partial}
    {\partial z}\right]
    +V^\sigma(z),
\end{equation}
where $\nabla^2_{xy}$ is the
2D Laplacian.
Since the structure of
Fig.\ref{structure} possesses translational invariance in the
x-y directions, the spin-dependent Kronig-Penney potential
$V^\sigma(z)$ and the effective mass $m^*(z)$ are functions of $z$ only.
Consequently the problem can be mapped onto a $k_\parallel$-dependent
1D problem, whose Hamiltonian is
\begin{equation} \label{kph1D}
    H^\sigma(z;k_\parallel)=-\frac{\hbar^2}{2}
    \frac{\mathrm d}{{\mathrm d} z}\frac{1}{m^*(z)}
    \frac{\mathrm d}{{\mathrm d} z}
    +\frac{\hbar^2k_\parallel^2}{2m^*(z)}+V^\sigma(z).
\end{equation}
For each $k_\parallel$ and spin $\sigma$, an eigenstate at the
Fermi energy contributes $e^2/h$ to the
conductance of this infinite periodic structure.
In the general case, the eigenstates can be obtained numerically using standard
transfer matrix techniques.
First consider the case of constant $m^*(z)$,
where the problem can be solved analytically.
Since the Hamiltonian
(\ref{kph1D}) depends  on $k_\parallel$ only through an energy shift,
one finds that the conductance per unit area has the simple form
\begin{equation} \label{cond}
    \Gamma=\frac{8\pi e^2 m^*}{h^3}\Delta=
    \frac{8\pi e^2 m^*}{h^3}\sum_n\Delta_n,
\end{equation}
where $\Delta_n$ is the bandwidth of the n-th energy band of the
Hamiltonian
\begin{equation} \label{band}
    H^\sigma(z)=-\frac{\hbar^2}{2m^*}\frac{{\mathrm d}^2}{{\mathrm d} z^2}
    +V^\sigma(z),
\end{equation}
and the sum is over all  occupied minibands.
Consider an infinite superlattice composed of materials A and B,
with layer-thicknesses $l_{\mathrm A}$ and $l_{\mathrm B}$
($l_{\mathrm A}+l_{\mathrm B}=L$), and  Kronig-Penney potential
 $V=V_o$ ($E_{\mathrm F}>V_o$) in the metal A and $V=0$ in
the metal B. If $k_\perp$ is the Bloch vector in the direction of the current,
the secular equation is
\begin{eqnarray}
    \cos(k_\perp L)=\cos(k_{\mathrm A}l_{\mathrm A}+k_{\mathrm B}l_{\mathrm B})- \label{secular} \\
    -\frac{(k_{\mathrm A}+k_{\mathrm B})^2}{k_{\mathrm A}k_{\mathrm B}} 
    \sin(k_{\mathrm A}l_{\mathrm A})\sin(k_{\mathrm B}l_{\mathrm B}), \nonumber
\end{eqnarray}
with $k_{\mathrm A}(E)=\sqrt{2m^*(E-V_o)}/\hbar$ and 
$k_{\mathrm B}(E)=\sqrt{2m^*E}/\hbar$. 
Based on this expression, we now argue that the bandwidths
exhibit several periods of oscillation
as the layer thicknesses are varied.

\begin{figure}
\narrowtext
\epsfysize=6cm
\epsfxsize=7cm
\centerline{\epsffile{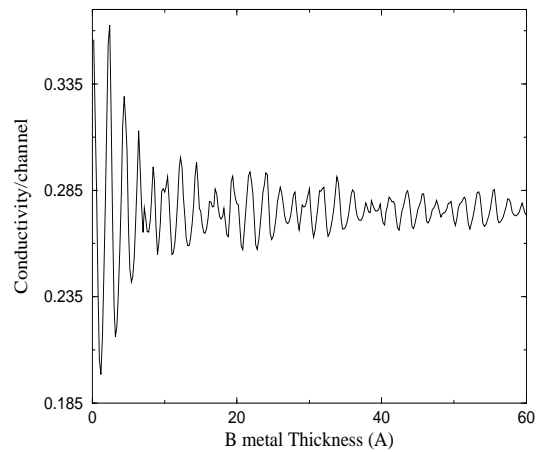}}
\caption{ \label{Fig4}
Conductivity per channel in the effective mass approximation. 
The parameters are $E_{\mathrm F}=10$eV, $V_o=6$eV, $m^*=0.5$MeV, $l_{\mathrm A}=8$\AA.}
\end{figure}

To describe Ni/Co multilayers, we vary the
thickness of metal B
keeping fixed the thickness of metal A.
To understand the oscillatory behaviour of the band-widths,
we note that  equation (\ref{secular}) cannot be satisfied
at energies for which
\begin{equation} \label{bandgap}
    k_{\mathrm A}(E)l_{\mathrm A}+k_{\mathrm B}(E)l_{\mathrm B}=m\pi,
\end{equation}
where $m$ is an integer.
Hence at $E=E_F$ and fixed $l_A$, successive bandgaps appear at
the Fermi energy $E_F$ when  $l_{\mathrm B}$ changes by
\begin{equation} \label{pfermi}
    l_{\mathrm B}^{m}=\frac{\pi}{k_{\mathrm B}(E_{\mathrm F})}m=
    \frac{\pi\hbar}{\sqrt{2m^*E_{\mathrm F}}}m=l_{\mathrm B}^{\mathrm F}m.
\end{equation}
Equation (\ref{pfermi}) introduces the first  period of oscillation
$l_{\mathrm B}^{\mathrm F}$. The second period corresponds to
the presence of narrow gaps below the Fermi energy.
From equations (\ref{secular}) and (\ref{bandgap})
narrow bandgaps appear at the energies
\begin{equation} \label{engap}
    E_A^{(n)}=\frac{\hbar^2\pi^2n^2}{2m^*l^2_{\mathrm A}}+V_o,
\end{equation}
whenever the lengths $l_{\mathrm B}$ equal
\begin{equation} \label{enper}
    l_{\mathrm B}^{(n)}=\frac{\pi\hbar}{\sqrt{2m^*E_A^{(n)}}}.
\end{equation}
The total bandwidth $\Delta$ and hence the conductance per unit of area
(\ref{cond}) are oscillating functions with  periods
$l_{\mathrm B}^{\mathrm F}$ and the $l_{\mathrm B}^{(n)}$'s.
All these periods are of order $\lambda_F$ (ie few \AA), but beating
between them can give rise to long-period oscillations. It is important to note
that the Fermi period is defined only through the Fermi energy, while the
periods $l_{\mathrm B}^{(n)}$ depend critically on the superlattice geometry.
In particular, because the energies corresponding to periods (\ref{engap})
depend on $1/l^2_{\mathrm A}$ and must not exceed the Fermi energy, the number 
of $l_{\mathrm B}^{(n)}$'s depends on the thickness of the metal A.
If $l_{\mathrm A}$ is large, a large number of $l_{\mathrm B}^{(n)}$ periods
will be present and the beating pattern will be complex.
On the other hand, if
$l_{\mathrm A}$ is small, few $l_{\mathrm B}^{(n)}$'s will be present, giving
rise to a simple beating pattern.
A numerical evaluation of Eq. (\ref{cond}) is shown in figure \ref{Fig4}.
For the chosen parameter in this plot, we expect
only one $l_{\mathrm B}^{(n)}$ and clear
beats are observed, with period $2l_{\mathrm B}^{(1)}l_{\mathrm B}^{\mathrm F}
/(l_{\mathrm B}^{(1)}-l_{\mathrm B}^{\mathrm F})$.
Since the $l_{\mathrm B}^{(n)}$ periods are
characteristic of the superlattice structure we predict that the period
of the long oscillations can be set by choosing the appropriate 
superlattice geometry. 

The above dependence of oscillations on the multilayer
structure is missed by a trilayer quantum
well approach to conductance oscillations and  GMR \cite{mat},
where only two periods have been identified. 
The first of these $p^{\mathrm FS}$ depends 
on the extremal Fermi surface radius of the spacer forming the well, and in the parabolic
band approximation corresponds exactly to the period $l_{\mathrm B}^{\mathrm F}$.
The second period $p^{\mathrm cp}$ depends on the cut-off of the sum over the $k_\parallel$'s,
and in the parabolic approximation, on the energy difference between the Fermi
energy and
the step potential $V_o$. In our superlattice description, this period is replaced
by the class of periods $l_{\mathrm B}^{(n)}$, which are a function of
the superlattice structure itself. This structural
dependence of the oscillation periods is the key to understanding the apparent
non-reproducibility of the long period oscillations from sample to sample, observed
in some of the experiments \cite{12bis}.
It may be shown that these beating features are preserved when a more realistic
material-dependent effective mass is used. Futhermore, for those
cases where we expect the mass to be significantly different in the two
materials (e.g. mainly s-p like in one material and mainly d-like in the other)
we have shown that the Kronig-Penney model reproduces the main features of
the more accurate tight-binding model, with physically reasonable
choices of the band offsets and effective masses.

Bearing in mind that our analysis describes  the CPP configuration,
we can also speculate on the absence of the oscillations in other recent
experiments \cite{12bis,kim}. Ag/Cu \cite{12bis} exhibits very good phase separation 
between
the different metals and hence it should be a good candidate for observing
conductance oscillations. However the band match between Ag and Cu is
very good, resulting in a very small scattering potential at the interface.
In the effective mass approach this means a very small step potential $V_o$
with respect to the Fermi energy. A large number of periods $l_{\mathrm B}^{(n)}$
will be present and the beats will be difficult to detect. The same argument
is valid for Ag/Au \cite{12bis}. In addition the high miscibility of
Ag and Au results in dirty interfaces.
Ag/Pd \cite{kim} is in theory a good candidate to
show conductance oscillations because of the large mismatch
between the Ag and Pd bands.
Unfortunately interdiffusion at the interface is difficult to avoid and
the elastic mean free path will be quite short. Finally,
we observe that for Ni/Co \cite{gall1,gall2,12bis}, the majority band
reproduces roughly the situation of Ag/Cu, while the scattering in the minority
band is quite large. According to the effective mass model the minority band
will possess a low conductance with large oscillations, while
the conductance of the majority band will be large and the oscillations small.
This is precisely what we
obtain from in the material specific tight-binding calculations.

In summary, we have investigated the possibility of large long-period
oscillations in metallic superlattices in the ballistic regime.
With the current perpendicular to the plane
and superlattices with pseudorandom layer thicknesses,
the oscillations are predicted by accurate tight-binding calculations.
An effective mass analysis provides a qualitative understanding of
the nature of the oscillations and
highlights their dependence on the superlattice geometry.

{\bf Acknowledgments}: This work is supported by the EPSRC and the EU
TMR Programme.

\vspace{1in}

\end{multicols}
\end{document}